\documentclass[prl,twocolumn,floatfix,amsmath,amssymb,superscriptaddress]{revtex4}
\usepackage{graphicx, subfigure}
\usepackage{dcolumn}
\usepackage{color}
\usepackage{bm}
\usepackage[T1]{fontenc}
\usepackage[matrix,frame,arrow]{xypic}

\newcommand{\ket}[1]{\left|#1\right>}
\newcommand{\bra}[1]{\left<#1\right|}
\newcommand{\tr}[1]{\text{Tr}\left(#1\right)}

\newcommand{\ba}{\text{\bf{a}}}

\newcommand{\bH}{\text{\bf{H}}}
\newcommand{\bD}{\text{\bf{D}}}
\newcommand{\bO}{\text{\bf{O}}}
\newcommand{\bI}{\text{\bf{I}}}
\newcommand{\brho}{\text{\boldsymbol{$\rho$}}}

\newcommand{\bsigma}{\text{\boldsymbol{$\sigma$}}}

\begin{document}

\title{Stabilizing a Bell state of two superconducting qubits by dissipation engineering}
\author{Z. Leghtas}
\affiliation{Departments of Applied Physics and Physics, Yale University, New Haven, Connecticut 06520, USA}
\affiliation{INRIA Paris-Rocquencourt, Domaine de Voluceau, B.P.~105, 78153 Le Chesnay Cedex, France}
\author{U. Vool}
\affiliation{Departments of Applied Physics and Physics, Yale University, New Haven, Connecticut 06520, USA}
\author{S. Shankar}
\affiliation{Departments of Applied Physics and Physics, Yale University, New Haven, Connecticut 06520, USA}
\author{M. Hatridge}
\affiliation{Departments of Applied Physics and Physics, Yale University, New Haven, Connecticut 06520, USA}
\author{S.M. Girvin}
\affiliation{Departments of Applied Physics and Physics, Yale University, New Haven, Connecticut 06520, USA}
\author{M.H. Devoret}
\affiliation{Departments of Applied Physics and Physics, Yale University, New Haven, Connecticut 06520, USA}
\author{M. Mirrahimi}
\affiliation{Departments of Applied Physics and Physics, Yale University, New Haven, Connecticut 06520, USA}
\affiliation{INRIA Paris-Rocquencourt, Domaine de Voluceau, B.P.~105, 78153 Le Chesnay Cedex, France}

\date{\today}

\begin{abstract}
{We propose a dissipation engineering scheme that prepares and protects a maximally entangled state of a pair of superconducting qubits. This is done by off-resonantly coupling the two qubits  to a low-Q cavity mode playing the role of a dissipative reservoir. We engineer this coupling by applying six continuous-wave microwave drives with appropriate frequencies. The two qubits need not be identical. We show that our approach does not require any fine-tuning of the parameters and requires only that certain ratios between them be large. With currently achievable coherence times, simulations indicate that a Bell state can be maintained over arbitrary long times with fidelities above $94\%$.  Such performance leads to a significant violation of Bell's inequality (CHSH correlation larger than 2.6) for arbitrary long times.}
\end{abstract}

\maketitle

Entanglement is a fundamental, yet counter-intuitive concept in
quantum mechanics. Maximally entangled two-qubit states, often called Bell
states, violate classical correlation properties~\cite{Bell-1964,CHSH-1969,Aspect-al-PRL_1981,Tittel-al-Gisin-PRL_1998} and are an essential building block
for quantum communication and quantum information. Unfortunately, these states are also difficult to
generate and sustain as interaction with the environment typically leads to
rapid loss of their unique quantum properties.
Therefore, stabilizing a Bell state is a sought after goal.

Quantum state stabilization can be achieved by an active feedback
loop in which the system is measured, and conditioned on the measurement of an error,  a gate restores the system to the desired
state \cite{Sayrin-al-Haroche-Nature_2011,Vijay_2012}.
This approach suffers from the latency of data acquisition and
analysis that must take place during the lifetime of the quantum system, a particularly acute problem for relatively short-lived quantum systems such as
superconducting qubits.
An alternative approach to stabilization is quantum {dissipation (reservoir)} engineering~\cite{Poyatos96}. In this approach, autonomous feedback is achieved  by drive-generated couplings between the quantum system and a dissipative reservoir \cite{plenio-et-al-1999,kraus-cirac-2004,paternostro-et-al-2004,mabuchi-et-al-PRL2010,kastoryano-et-al-2011,Barreiro-et-al:Nature_2011,Krauter2011,stannigel-et-al-2012,PRancestor}. Thus, this type of autonomous feedback corrects errors without 
any need for real-time processing of the measured record. 

{In the field of quantum Josephson circuits, autonomous stabilization of single qubit
states has recently been implemented by using a low-Q cavity mode as a dissipative reservoir~\cite{murch-et-al-2012,Geerlings-et-al_2012}. { While schemes for autonomous stabilization of entangled states exist in atomic physics contexts \cite{plenio-et-al-1999,paternostro-et-al-2004,kastoryano-et-al-2011,stannigel-et-al-2012}, the required symmetry in the physical system would appear to be an obstacle to implementing these with superconducting qubits.
Indeed, while qubits in atomic physics experiments naturally possess identical frequencies and coupling strengths to an engineered
environment, such symmetry is difficult to achieve with superconducting
artificial atoms. 
On the other hand, since these systems are engineered, they benefit from a large design versatility. In circuits, a large set of Hamiltonians can be designed for dissipation engineering schemes. Moreover, coupling strengths can be tuned by simple modifications of circuit elements, and hence are not fixed by fundamental constants, as in the case of atomic physics systems.

In this Letter, we propose to stabilize a maximally entangled
state 
of a pair of superconducting qubits. Since we require no strong built-in symmetry in the physical setup, the proposed method is readily implementable
with superconducting qubits  and should lead to breakthrough performance with currently achievable experimental
parameters. The scheme exploits the strong dispersive
interaction~\cite{schuster-nature07} between two superconducting qubits and a single
low-Q cavity mode and only requires continuous-wave (CW) microwave drives with well-chosen fixed frequencies, amplitudes and phases (see Fig.~\ref{fig:experimentalsetup}). Moreover, our scheme is robust against small variations of the latter control parameters, and requires only some basic calibration. By avoiding resonant interactions
between the qubits and the cavity mode the qubits remain protected against the Purcell
effect which would reduce their coherence times.

Here, the low-Q cavity mode acts as an engineered reservoir which evacuates entropy from the qubits when a perturbation occurs: by driving the qubits and cavity with CW drives, we induce an autonomous feedback loop which corrects the state of the qubits every time it decays out of the desired Bell state. In order to show the robustness of our scheme and its ability to prepare and protect a state significantly violating Bell's inequality, we have performed simulations with experimentally achievable parameters and realistic decoherence values.

\begin{figure}
\includegraphics[width=\columnwidth]{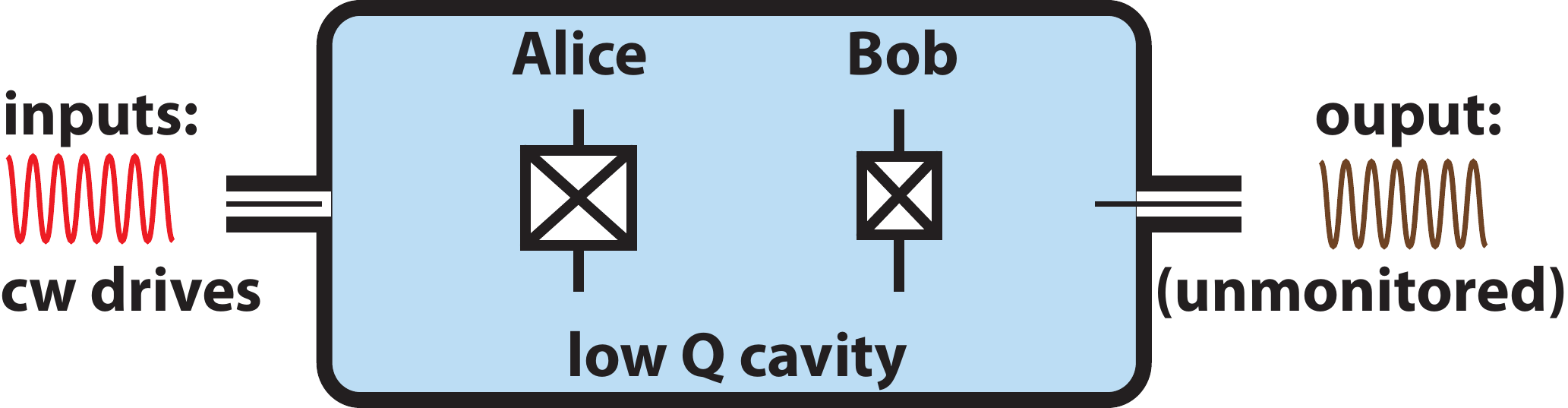}
\caption{Possible setup for the proposed experiment: two superconducting qubits (Alice and Bob) interact with a mode of a 3D cavity. CW tones are applied on the input port of the cavity. These tones drive the qubits and the cavity and stabilize a Bell state of the two qubits. This stabilization is autonomous: no analysis of the output signal is needed.}
\label{fig:experimentalsetup}
\end{figure}
Consider two non-identical qubits coupled to a
low-Q cavity mode (see Fig.~\ref{fig:experimentalsetup}). {The Hamiltonian of the three body system (two qubits and one cavity mode) in the Jaynes-Cummings approximation reads \cite{Haroche-Raimond_2006}:
\begin{multline*}
   \bH = \omega_{A}\frac{\bsigma_z^A}{2}+\omega_{B}\frac{\bsigma_z^B}{2}+\omega_c \ba^\dag\ba,\\
   + g_A (\bsigma_+^A\ba+\bsigma_-^A\ba^\dag)+g_B (\bsigma_+^B\ba+\bsigma_-^B\ba^\dag),
\end{multline*}
where $\bsigma_z^A$, $\bsigma_z^B$ are the Pauli $\bsigma_z$ operators for the Alice and Bob qubit respectively, and $\ba^\dag$($\ba$)  is the cavity creation (annihilation) operator. Here, $\omega_{A}$, $\omega_{B}$ and $\omega_c$ are the resonance frequencies for Alice, Bob  and the cavity mode respectively, and $g_A$, $g_B$ are the coupling
constants of the cavity to Alice and Bob respectively.  In the dispersive regime, $|\Delta_A|=|\omega_A-\omega_c|\gg g_A$ and $|\Delta_B|=|\omega_B-\omega_c|\gg g_B$,} and in the rotating frame for both qubits and the cavity, we obtain the following
effective Hamiltonian:
\begin{equation}
 {\bH_\text{eff}=(\chi_{A}\frac{\bsigma_z^A}{2}+\chi_{B}\frac{\bsigma_z^B}{2})
  \ba^\dag\ba\;},
\end{equation}
{where $\chi_A=2 {g_A}^2/\Delta_A$ and $\chi_B=2{g_B}^2/\Delta_B$ are the
dispersive coupling strengths between the qubits and the cavity mode.} We neglect the direct qubit-qubit interaction, which has no effect on our stabilization protocol. The same effective Hamiltonian would be obtained by more realistic quantum circuit models \cite{nigg-et-al-2012}. Here, we
consider the coupling to be in the 
strong dispersive
regime~\cite{schuster-nature07}, where $\chi_{A}$ and $\chi_B$ are assumed to be larger than the cavity and qubit line-widths $\chi_{A},\chi_B>\kappa \gg 1/T_{2}^A, 1/T_{2}^B$. In this regime, the resonance frequency of each qubit depends on the number of photons in the cavity, and conversely the cavity frequency depends on the states of the qubits (see Fig.~\ref{fig:energyleveldiagram}~[a,c]).
We can, hence, drive the qubits (cavity mode) conditionally on the state of
the cavity mode (qubits). The stabilization drives lead to a measurement induced dephasing at a rate $\kappa_m$ which increases with the difference of the dispersive shifts $|\chi_A-\chi_B|$. Thus, the only symmetry requirement for our dissipation engineering scheme is that $\chi_A,~\chi_B$ are sufficiently close to each other to ensure $\kappa_m\ll \kappa$, since $\kappa$ is the time scale associated with the autonomous feedback loop. Using \cite{Gambetta-al-PRA_2006}, we show that this condition is satisfied when 
\begin{equation}
\label{eq:Xi1Xi2}
|\chi_A-\chi_B|\ll \chi_A\chi_B/\kappa\sqrt{\bar n}\;.
\end{equation}
Note that such symmetry is easily achieved with superconducting qubits, by tuning the frequency of one of the qubits so that $2{g_A}^2/\Delta_A$ and $2{g_B}^2/\Delta_B$ become close enough. 

We denote as $\omega_c^{gg}, \omega_c^{ge}, \omega_c^{eg}$ and $\omega_c^{ee}$ the cavity frequencies when the qubits are respectively in the states $\ket{gg},
\ket{ge}, \ket{eg}$ and $\ket{ee}$. We have used the notation
$\ket{ge}=\ket{g}\otimes\ket{e}$, where the first (second) element refers to the
state of the Alice (Bob) qubit. Similarly we denote $\omega_{A}^{n}$
($\omega_{B}^{n}$) the frequency of Alice (Bob) when there are $n$
photons in the cavity. 

\begin{figure}
\includegraphics[width=\columnwidth]{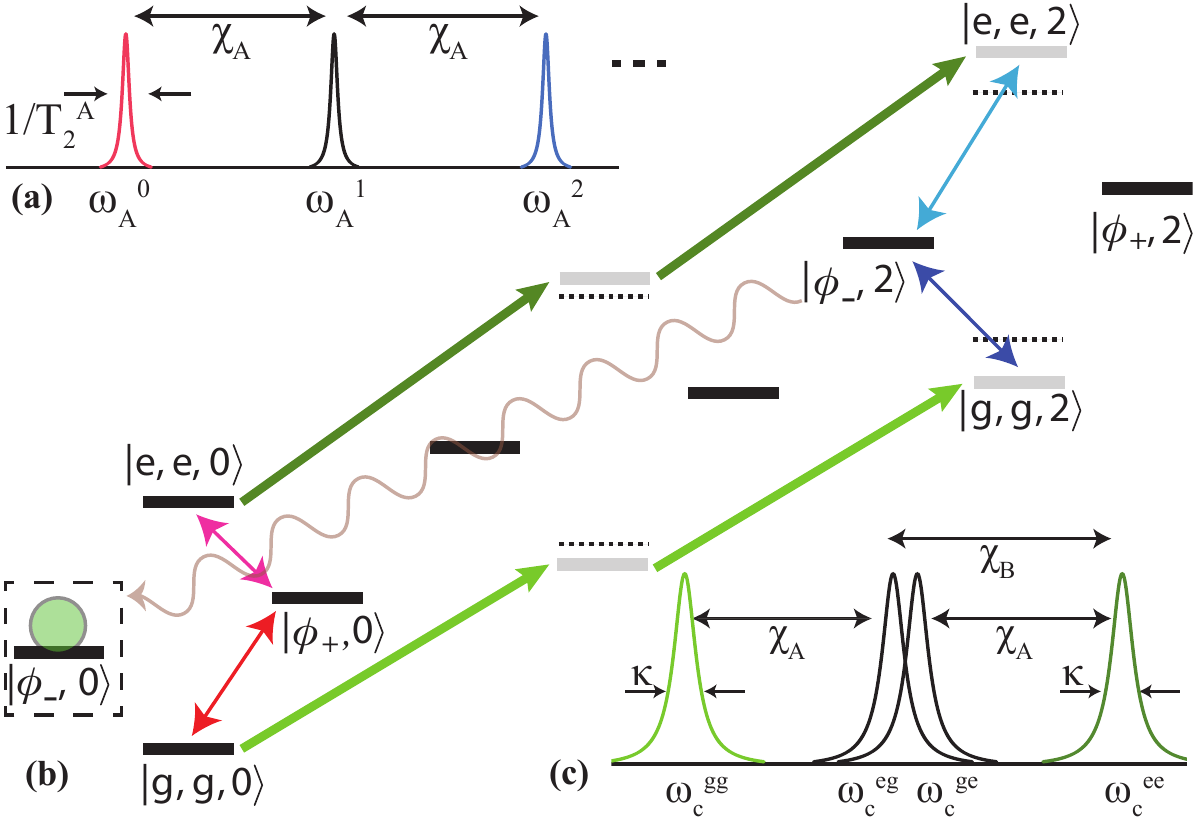}
\caption{ (a), (c) Spectra of qubit Alice and the cavity, respectively. The peaks are resolved since $\chi_{A},\chi_B > \kappa,1/T_2^A,1/T_2^B$. (b) Energy level diagram of two qubits off-resonantly coupled to a cavity mode. Straight line arrows indicate couplings between energy levels which are induced by CW drives. The wavy arrow indicates energy decay from the cavity mode. These drives couple the two qubits to the cavity, here used as a quantum reservoir, in such a way that the Bell state $\ket{\phi_-}=(\ket{ge}-\ket{eg})/\sqrt{2}$ is the only stable state.}
\label{fig:energyleveldiagram}
\end{figure}

{We now detail, step by step, the dissipation engineering procedure by explaining the task performed by each CW drive (see Fig.~\ref{fig:energyleveldiagram}).}

{\textbf{Parity selection:} Applying two CW drives on the resonator at frequencies $\omega_c^{gg}$ and $\omega_c^{ee}$ with equal amplitudes $\epsilon_c=\frac{\kappa}{2}\sqrt{\bar n}$, we generate a coherent state with mean photon number $\bar n$ in the cavity when the qubits are in $\ket{gg}$ or $\ket{ee}$. Hence, an initial state $\ket{gg,0}$ ($\ket{ee,0}$) will converge to a state $\ket{gg,\alpha e^{i\frac{\chi_A+\chi_B}{2} t}}$ ($\ket{ee,\alpha e^{-i\frac{\chi_A+\chi_B}{2} t}}$) where  $\ket{\alpha e^{i\phi}}$ is a coherent state with  $|\alpha|^2=\bar n$. On the other hand, states of the form $c_{eg}\ket{eg,0}+c_{ge}\ket{ge,0}$ are left invariant. Indeed, we assume $|\omega_c^{eg,ge}-\omega_c^{ee,gg}|=\chi_A, \chi_B\gg \epsilon_c=\frac{\kappa}{2}\sqrt{\bar n}$, and hence the drives are off-resonant with the cavity when the qubits are in an odd parity state, leaving the cavity in vacuum. Note that for finite ratios $\chi_{A,B}/\epsilon_c$, the cavity will populate slightly when the qubits are in an odd parity state, inducing a dephasing $\kappa_m$ which will be much smaller than the correction rate $\kappa$ as long as condition \eqref{eq:Xi1Xi2} is satisfied. At this stage, if we measure the photons leaking out of the cavity, we can determine if the state of the qubits is $\ket{gg}, \ket{ee}$ or in the odd parity manifold span$\{\ket{ge}, \ket{eg}\}$.}


{\textbf{Bell state selection:} We now  apply two Rabi drives  separating the two Bell states $\ket{\phi_-}=(\ket{ge}-\ket{eg})/\sqrt{2}$ and $\ket{\phi_+}=(\ket{ge}+\ket{eg})/\sqrt{2}$ in the odd parity manifold. Since the cavity mode remains in the vacuum state within this manifold, we apply two qubit drives at frequencies $\omega_{A}^0$ and $\omega_{B}^0$, and furthermore the corresponding Rabi amplitudes are chosen to be the same and given by $\Omega^0$. Assuming with no loss of generality that these drives are around the X axis, this adds the term $\Omega^0(\bsigma_x^A+\bsigma_x^B)$ to the Hamiltonian in the rotating frame of the qubits. The latter term leaves invariant the state $\ket{\phi_-}$, and couples the state $\ket{\phi_+}$ to $(\ket{gg}+\ket{ee})/\sqrt{2}$. These two qubit drives, added to the two cavity drives, pump the population from $\ket{\phi_+,0}$ to a mixture of the states $\ket{gg,\alpha e^{i\frac{\chi_A+\chi_B}{2} t}}$ and $\ket{ee,\alpha e^{-i\frac{\chi_A+\chi_B}{2} t}}$, while keeping the state $\ket{\phi_-,0}$ invariant.  Hence, by looking at the photons leaking out of the cavity, we can determine if the qubits are in the desired target state $\ket{\phi_-}$, {or if they have been projected to $\ket{gg}$ or $\ket{ee}$}. }

\textbf{Irreversible pumping to the target Bell state:} To complete the dissipation engineering scheme which stabilizes $\ket{\phi_-}$, we need to pump the population from $\ket{gg,\alpha e^{i\frac{\chi_A+\chi_B}{2} t}}$ and $\ket{ee,\alpha e^{-i\frac{\chi_A+\chi_B}{2} t}}$ back to $\ket{\phi_-}$. To this end, we apply two more drives at frequencies $\omega_{A}^{0}+\bar n \frac{\chi_A+\chi_B}{2}$ and $\omega_{B}^{0}+\bar n \frac{\chi_A+\chi_B}{2}$, with the same Rabi frequency $\Omega^{\bar n}$. Adjusting the phase difference of these drives appropriately, we obtain the following term in the Hamiltonian (in the rotating frame of the qubits): $\Omega^{\bar n}\left(e^{-i\bar n\frac{\chi_A+\chi_B}{2}t}(\bsigma_+^A-\bsigma_+^B)+\text{c.c.}\right)$ (c.c. stands for complex conjugate). For integer $n\approx\bar n$, these drives couple the states {$\ket{gg,n}$ and $\ket{ee,n}$}  to $\ket{\phi_-,n}$. The latter state is unaffected by the cavity drives and so decays down to $\ket{\phi_-,0}$, as indicated by Fig.~\ref{fig:energyleveldiagram}.

{In summary, turning on the above six CW drives simultaneously  pumps the population from any of the states $\ket{gg}$, $\ket{ee}$ and $\ket{\phi_+}$ to the state $\ket{\phi_-}$. This pumping happens at a rate of order $\kappa$: the states $\ket{gg,0}$ and $\ket{ee,0}$ converge to the states $\ket{gg,\alpha e^{i\frac{\chi_A+\chi_B}{2} t}}$ and $\ket{ee,\alpha e^{-i\frac{\chi_A+\chi_B}{2} t}}$ at  rate $\kappa$; the state $\ket{\phi_-,n}$ decays towards $\ket{\phi_-,0}$ at rate $\kappa$; and we can choose the oscillation rates $\Omega^0$ and $\Omega^{n}$ at will. Provided that this pumping rate of order $\kappa$ is much larger than the decoherence rates of the qubits that destroy entanglement, the dissipation engineering efficiently prepares and protects the desired Bell state. Note, furthermore, that any imperfection in the experimental parameters, such as non-ideal tuning of the drives and a difference between $\chi_A$ and $\chi_B$ can also lead to loss of coherence for both qubits: however, as long as the induced dephasing rate is smaller than the pumping rate, the protocol remains efficient.}

\begin{figure}
\includegraphics[width=\columnwidth]{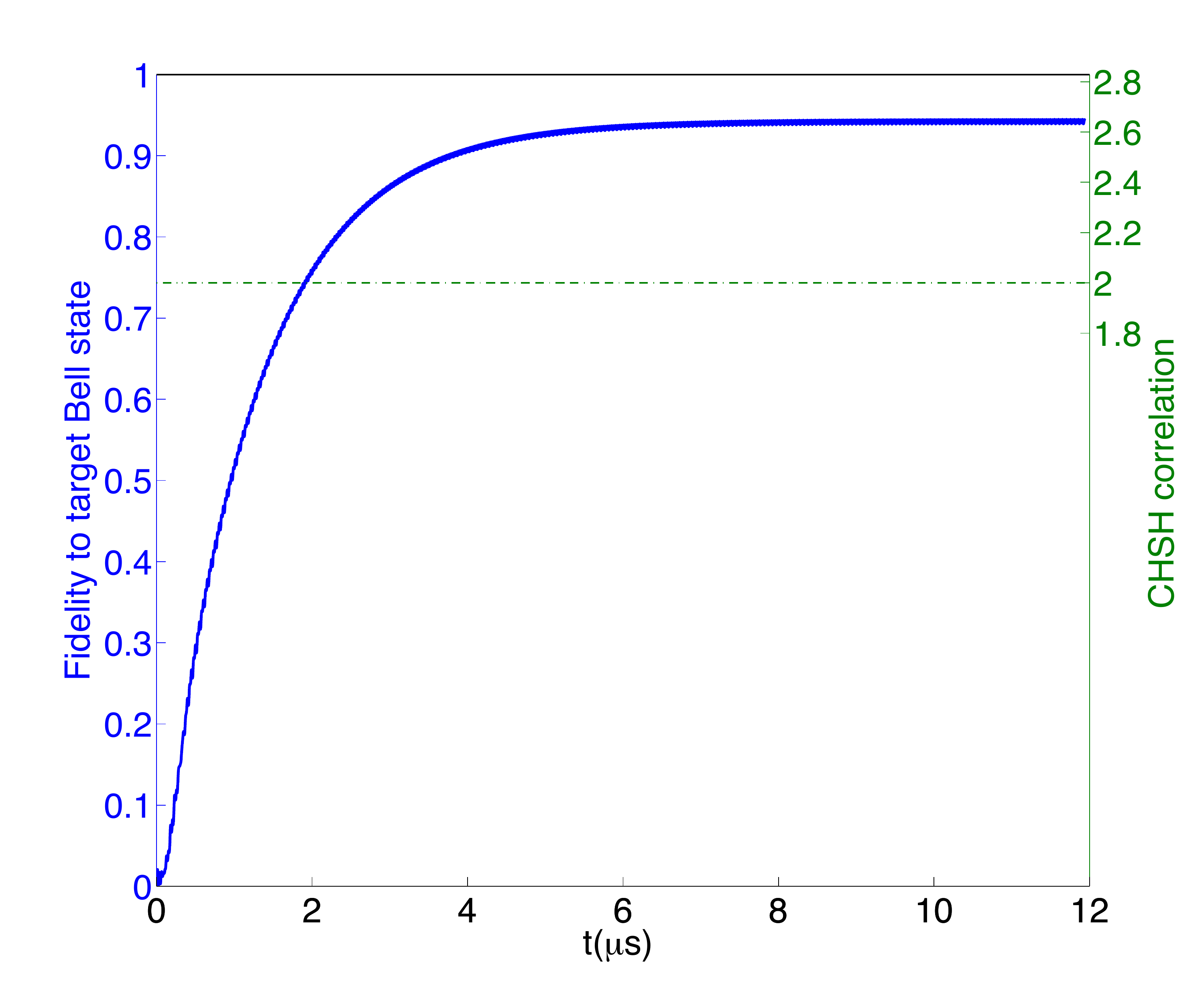}
\caption{Simulation of the fidelity with respect to the target state $\ket{\phi_-}$ as a function of time. The CHSH correlation associated to each fidelity can be read on the right axis. The system parameters
are $T_{1,2}^{A,B}=50~\mu$s, $\kappa/2\pi=2~$MHz, $\chi_A/2\pi=10~$MHz,
$\chi_B/2\pi=9.5~$MHz. The cavity is driven {with $\bar{n}=4$, the qubit drive
strengths are $\Omega^0=\frac{\kappa}{2}$ and $\Omega^{\bar{n}}=\kappa$} and the
initial state is the ground state $\ket{gg,0}$. Notice that realistic decoherence mechanisms and imperfect parameter tuning are
included in the simulation. {Despite these imperfections, the desired Bell state is stabilized with a fidelity of about $94\%$ resulting in a constant violation of Bell's inequality
(stable CHSH correlation of 2.64 close to the maximal violation limit $2\sqrt 2\approx 2.83$).}}
\label{fig:FidelityBell_vs_time}
\end{figure}

\begin{figure}
\includegraphics[width=\columnwidth]{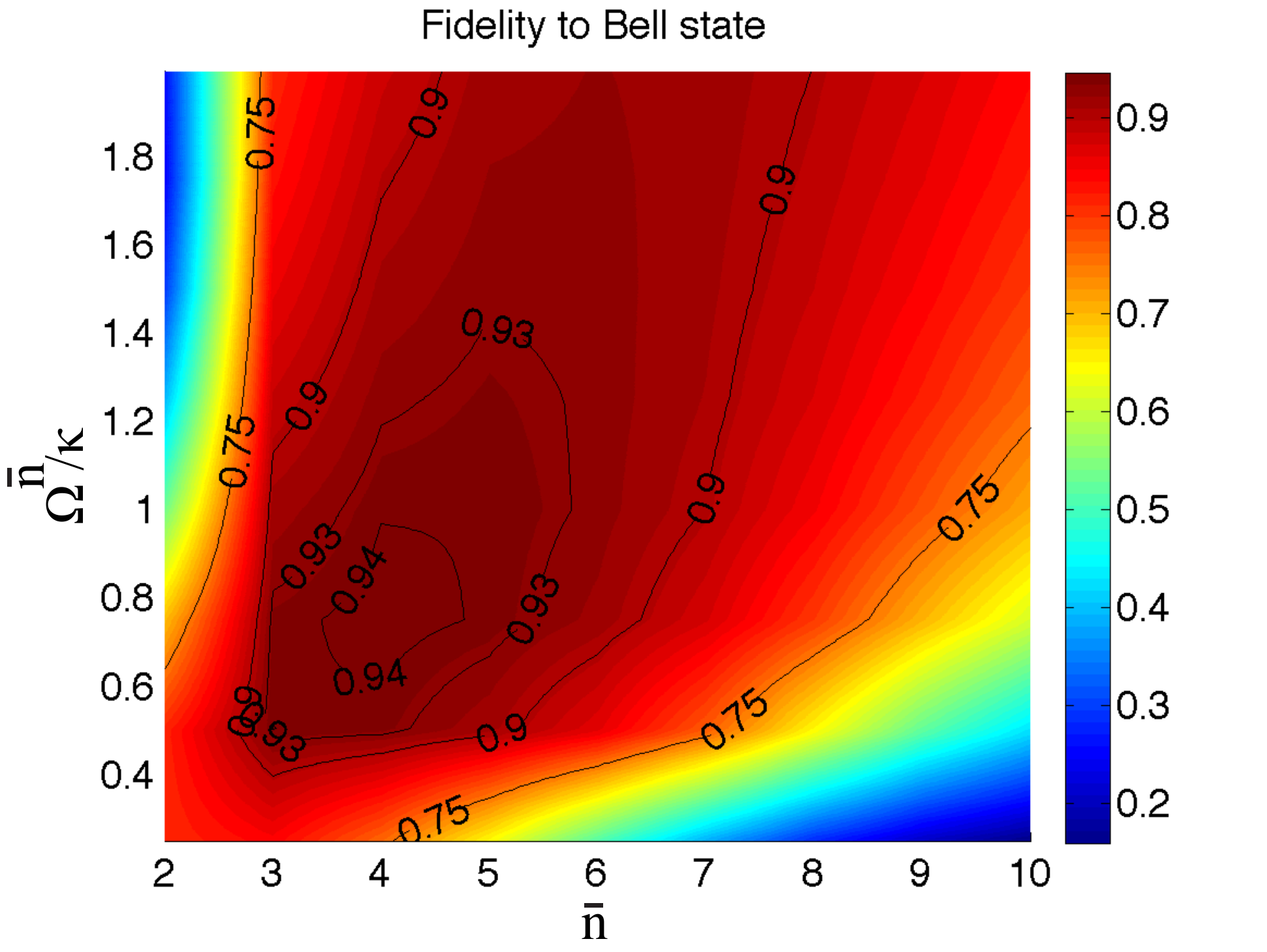}
\caption{Fidelity (color) of the two qubit steady state with respect to the pure target state $\ket{\phi_-}$, as a function of the two drive strengths. The size of the 90\% fidelity contour demonstrates the robustness of the protocol. The parameter domain where a Bell violation would occur is even larger (75\% contour). Large $\bar n$ contradicts the constraint \eqref{eq:Xi1Xi2}, thus leading to a higher induced dephasing between the two qubits. On the other hand, for small $\bar n$, we are constrained to a small range of $\Omega^{\bar n}$ in order to satisfy $\bar n\chi_{A,B} \gg \Omega^{\bar n}$, which is needed to
perform conditional qubit operations. Taking the parameters of the previous simulations, the optimal compromise is around $3\le\bar n\le5$ and $0.5\le\Omega^{\bar n}/\kappa\le1$, leading to a fidelity greater than 93\%.}
\label{fig:Bell_vs_nbar_Omeganbar}
\end{figure}

We present simulations of the Lindblad master equation with practically
achievable parameters to address all these imperfections and demonstrate our scheme is
robust to them. The system parameters we take for the simulation are $T_{1,2}^{A,B}=50~\mu$s, $\kappa/2\pi=2~$MHz, $\chi_A/2\pi=10~$MHz, $\chi_B/2\pi=9.5~$MHz. For transmon qubits, a Purcell filter \cite{Reed-al-APL_2010} might be necessary to obtain $\kappa T_{1,2}^{A,B}\gg1$.

 We simulate the Lindblad master equation
 \begin{eqnarray*}
 	\frac{d}{dt}\brho(t)&=&-\frac{i}{\hbar}[\bH(t),\brho(t)]+\kappa\bD[\ba]\brho(t) \\
 	&+&\sum_{j=A,B}\left({\frac{1}{
 	T_1^{j}}\bD[\bsigma_-^{j}]\brho(t)+\frac{1}{2 T_{\phi}^{j}}\bD[\bsigma_z^{j}]\brho(t)}\right)\;\notag,
 \end{eqnarray*}
 where
 \begin{eqnarray*}
 \bH(t)&=&(\chi_A\frac{\bsigma_z^A}{2}+\chi_B\frac{\bsigma_z^B}{2}) \ba^\dag\ba\\
 &+&2\epsilon_c\cos(\frac{\chi_A+\chi_B}{2} t)(\ba+\ba^\dag)+\Omega^0(\bsigma_x^A+\bsigma_x^B)\\
 &+&\Omega^{\bar n}\left(e^{-i\bar n\frac{\chi_A+\chi_B}{2}t}(\bsigma_+^A-\bsigma_+^B)+\text{c.c.}\right),
 \end{eqnarray*}
 with $\bsigma_+=(\bsigma_-)^\dag=\ket{e}\bra{g}$. Furthermore, $1/T_\phi^{A,B}=1/T_2^{A,B}-1/2T_1^{A,B}$ are the pure dephasing rates and the Lindblad super-operator is defined for any observable $\bO$ as
 \begin{equation*}
 \bD[\bO]\brho=\bO\brho\bO^\dag-\frac{1}{2}\bO^\dag\bO\brho-\frac{1}{2}\brho\bO^\dag\bO\;.
 \end{equation*}
{Choosing optimal drive amplitudes, we present the evolution of our system
in} Fig.~\ref{fig:FidelityBell_vs_time}.
In this figure we plot the fidelity $F(t)$ of the two qubit state with respect to the
pure target state $\ket{\phi_-}$. The corresponding CHSH correlation $B(t)$ can be read on the right hand side axis. More precisely, we have $F(t)=\tr{(\ket{\phi_-}\bra{\phi_-}\otimes \bI_c) \brho(t)}$ and $B(t)=\tr{(\bO_\text{CHSH}\otimes \bI_c)\brho(t)}$, where 
\begin{align*}
\bO_\text{CHSH}=\bsigma_y^A\frac{-\bsigma_y^B-\bsigma_x^B}{\sqrt{2}}+\bsigma_x^A\frac{-\bsigma_y^B-\bsigma_x^B}{\sqrt{2}}\\
+\bsigma_x^A\frac{\bsigma_y^B-\bsigma_x^B}{\sqrt{2}}-\bsigma_y^A\frac{\bsigma_y^B-\bsigma_x^B}{\sqrt{2}},
\end{align*}
and $\bI_c$ is the identity operator on the cavity.
 The state violates Bell's inequality when the
CHSH correlation rises above 2 (dash-dotted green horizontal line) \cite{NielsenChuang} .

{Fixing the experimental parameters, one can scan the drive amplitudes to find the maximal violation of the Bell inequality. This has been performed in Fig.~\ref{fig:Bell_vs_nbar_Omeganbar} where we scan $\bar n$ and $\Omega^{\bar n}$. As can be seen, for a large range of amplitudes, the Bell inequality remains violated which shows the robustness of our protocol to these control parameters. A discussion on the optimal values of the amplitudes is provided in the caption of the figure.} 

In conclusion, we have presented a scheme to stabilize a
Bell state without the need for active feedback, but rather by engineering a coupling to a
dissipative reservoir which performs an autonomous feedback loop. Our protocol is robust to system and control parameters, and does not require the two qubits to be identical. Numerical simulations taking into account all relevant experimental imperfections predict a CHSH correlation above 2.6 (higher than 94\% fidelity to the target Bell state) clearly violating Bell's inequality for arbitrarily long times. This Bell state could be further purified by monitoring the cavity output and selecting the state only when no photons are leaking out from the cavity. Preliminary experimental results are in excellent agreement with this theory and will be published elsewhere.

{
Facilities use was supported by the Yale Institute for
Nanoscience and Quantum Engineering (YINQE) and the
NSF MRSEC DMR 1119826. This research was supported in
part by the Office of the Director of National Intelligence
(ODNI), Intelligence Advanced Research Projects Activity
(IARPA), through the Army Research Office (W911NF-09-1-0369),
and in part by the U.S. Army Research Office (W911NF-09-1-0514).
M.M. and Z.L. acknowledge partial
support from the Agence National de Recherche under the
project EPOQ2, ANR-09-JCJC-0070. The authors  acknowledge
support from the NSF DMR 1004406}.

\bibliography{BellStateStabilization}

\end{document}